\RequirePackage{lineno}

\documentclass[aps,prd,onecolumn,superscriptaddress,showpacs,longbibliography]{revtex4-2}
\usepackage{graphicx}
\usepackage{dcolumn}
\usepackage{bm}

\begin{document}

\rightline{JLAB-PHY-23-3761}

\title{Motivation for Two Detectors at a Particle Physics Collider\\}

\author{Paul D. Grannis}
 \altaffiliation{Stony Brook University}
\author{Hugh E. Montgomery}
 \altaffiliation{Thomas Jefferson National Accelerator Facility}

\date{\today}

\begin{abstract}
It is generally accepted that it is preferable to build two general purpose detectors at any given collider facility. We reinforce this point by discussing a number of aspects and particular instances in which this has been important. The examples are taken mainly, but not exclusively, from experience at the Tevatron collider. 
\end{abstract}

\maketitle



\section{\label{sec:intro}Introduction}

The Electron Ion Collider (EIC)~\cite{eic} is a new facility being constructed 
at Brookhaven National Laboratory in collaboration with 
Thomas Jefferson National Accelerator Laboratory and their domestic and
international partners. In early 2022, the experimental program was 
discussed by a distinguished “blue Ribbon” panel of physicists. 
The committee was asked to adjudicate on the choice of a first detector 
based on the submission of different designs by three collaborations. 
While choosing one of the three, the committee also commented on the 
potential for improvement based on a substantial merging of the ideas, 
thus setting the path towards a strong single detector. 
The committee was also asked to consider the need for a second detector. 
It based its rather positive comments on the desire for complementarity 
between detector designs, the desire for the ability to confirm discoveries 
or results more generally, and eventually the potential to combine 
the results from the two experiments. The desirability of all 
these outcomes is based on reducing the risk associated with specific designs, 
allowing for different focus on physics questions, and providing a potential 
complementarity of the systematic and statistical uncertainties.

The funding perceived to be available for the construction of the 
accelerator and detectors is limited and thought to accommodate, with 
difficulty, a single intersection region and detector. Approval of 
a second detector has been deferred with the idea that the delay would 
facilitate the funding and the development of desired complementary technologies. 

There is a strong counter argument that it is highly desirable that 
the two detectors, even if starting at different times, are operating 
concurrently for a substantial part of the lifetime of the collider. 

Subsequent to the establishment of the first detector collaboration, 
Laboratory management, in conjunction with the EIC Users Group created 
a “2nd Detector Working Group”~\cite{eic-wg}. At a workshop~\cite{eic-wkshp} to kick-off the 
activities associated with the second detector, one of us (HEM) was 
invited to consolidate or amplify the case for the second detector based on recollections and specific experiences of similar situations in the past. 
This paper closely follows the discussion presented at that workshop 
but also contains some examples not included at that time. The two authors 
were members of the DØ collaboration which built and operated 
the “second” general purpose experiment at the Fermilab Tevatron.

In Section~\ref{sec:history}, we briefly tabulate 
the evolution of collider facilities and 
detectors since ~1980. In Section~\ref{sec:tevatron}, 
we provide a detailed discussion of 
examples from the two experiments, CDF and DØ, at the Tevatron. 
These cover instances where there were initial disagreements followed by 
confirmation, corrections of one experiment by the other, and eventual 
exploitation of the combined results of both to achieve enhanced sensitivity.  
In Section~\ref{sec:lhc}, we strengthen the interest in technological diversity 
and complementarity by looking at the two general purpose LHC experiments, 
ATLAS, and CMS.   Some concluding remarks are given in Section~\ref{sec:summary}.

\section{\label{sec:history}The Historical Norm}

For most of the “fixed target” era of particle physics, and for collider 
facilities through the 1970’s, any particular experiment did not constitute a 
significant addition to the accelerator investment, so individual experiments 
competed and cross checked each other. For collider facilities after about 1980, 
the individual experiments became more costly. Nevertheless, the desire for 
competition and verification motivated two or more general purpose experiments at 
each facility. With variations to accommodate examples where two similar colliders 
were involved, the following is historically how the situation developed. 

\begin{itemize}

\item	S${\rm p}\bar{\rm p}$S experiments: UA1, UA2
\item	$e^+e^-$experiments: SLC (Mark II, SLD), LEP (ALEPH, DELPHI, L3, OPAL)
\item	Tevatron experiments: CDF, DØ
\item	SSC experiments: GEM, SDC 
\item	HERA experiments: H1, ZEUS
\item	RHIC experiments: PHENIX, STAR
\item	B-Factory experiments: BaBar (at PEPII), BELLE (at KEKB) 
\item	LHC experiments: ATLAS, CMS
\end{itemize}

There are exceptions. BELLE II at Super-KEKB has no analogous direct 
competitor as had been provided by BaBar for BELLE.  LHCb, the specialized 
$B$-physics detector at the LHC, has no direct competitor with comparable 
scope but there are partial overlaps in physics capabilities with ATLAS and CMS 
on the one hand and with BELLE II on the other.

\section{\label{sec:tevatron}The Tevatron: CDF and D0}

\subsection{\label{sec:run1}Run 1, through 1996}

Both the S${\rm p}\bar{\rm p}$S~\cite{spps} and the Tevatron~\cite{tevatron} 
were derived from pre-existing infrastructure. Colliding protons with 
anti-protons obviated the need for a second collider ring, and 
stochastic cooling enabled the creation of beams of antiprotons of 
meaningful intensity. Starting with the ISR, the Tevatron was the 
third hadron collider and benefited, as did UA1 and UA2 at the 
S${\rm p}\bar{\rm p}$S collider, from lessons learned at the ISR, 
and from the electron-positron colliders, PETRA and PEP. 
In particular, the need for detectors with nearly 4$\pi$ coverage,
so as to enable a measurement of the missing transverse momentum 
carried by non-interacting particles, was generally accepted.

CDF~\cite{cdf}, the first Tevatron detector, was initiated by Fermilab 
in 1978 and  was designed in 1980 -- 1981. It detected a handful of 
collisions in 1985 and took its initial data between 1987 and 1989. Upgrades on 
a modest scale started even before the 1992 – 1996 Run 1, particularly a novel 
silicon vertex detector. A major upgrade took place in 1997 -- 2001 before 
operation in Run 2.

DØ~\cite{d0} was the second general purpose detector at the Tevatron. 
Several precursor proposals were rejected before the FNAL Director invited 
one of us (PDG) to pull together a collaboration to design an experiment 
already given preliminary approval.  The DØ detector was conceived during 
the period 1983 -- 1984, saw first collisions in 1992, and underwent a major 
upgrade in 1996 -- 2001.  Now in phase with CDF, DØ was active through 
Run 2 until the Tevatron ceased operations in 2011. 

\begin{figure}[hbt]
\includegraphics[width=0.6\textwidth,angle=0]{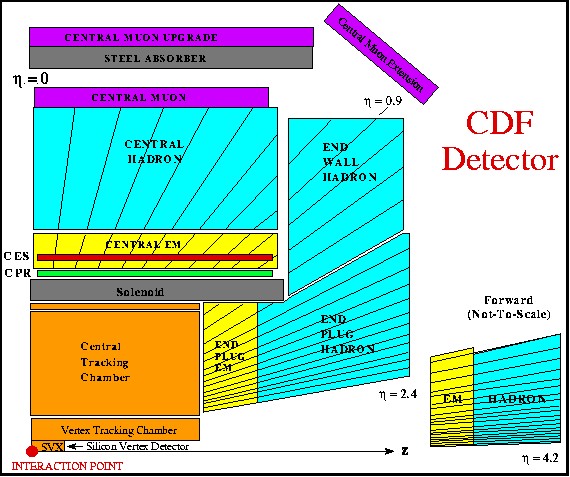}
\caption{\label{fig:CDF-detector} The CDF detector layout, circa 1992.}
\end{figure}

The initial CDF detector (Fig.~\ref{fig:CDF-detector}) was 
dominated by a large central superconducting 
solenoid containing a central tracker consisting initially of wire chambers that 
were soon supplemented by a vertex silicon detector. The barrel calorimeter, 
consisting of lead and iron/scintillator wedges, extended to $|\eta |$=1.  
Muon detection was accomplished by a number of separate systems. 
The silicon detector, installed before the 1992 running, consisted of
three silicon barrel layers.
           
\begin{figure}[hbt]
\includegraphics[width=0.65\textwidth,angle=0]{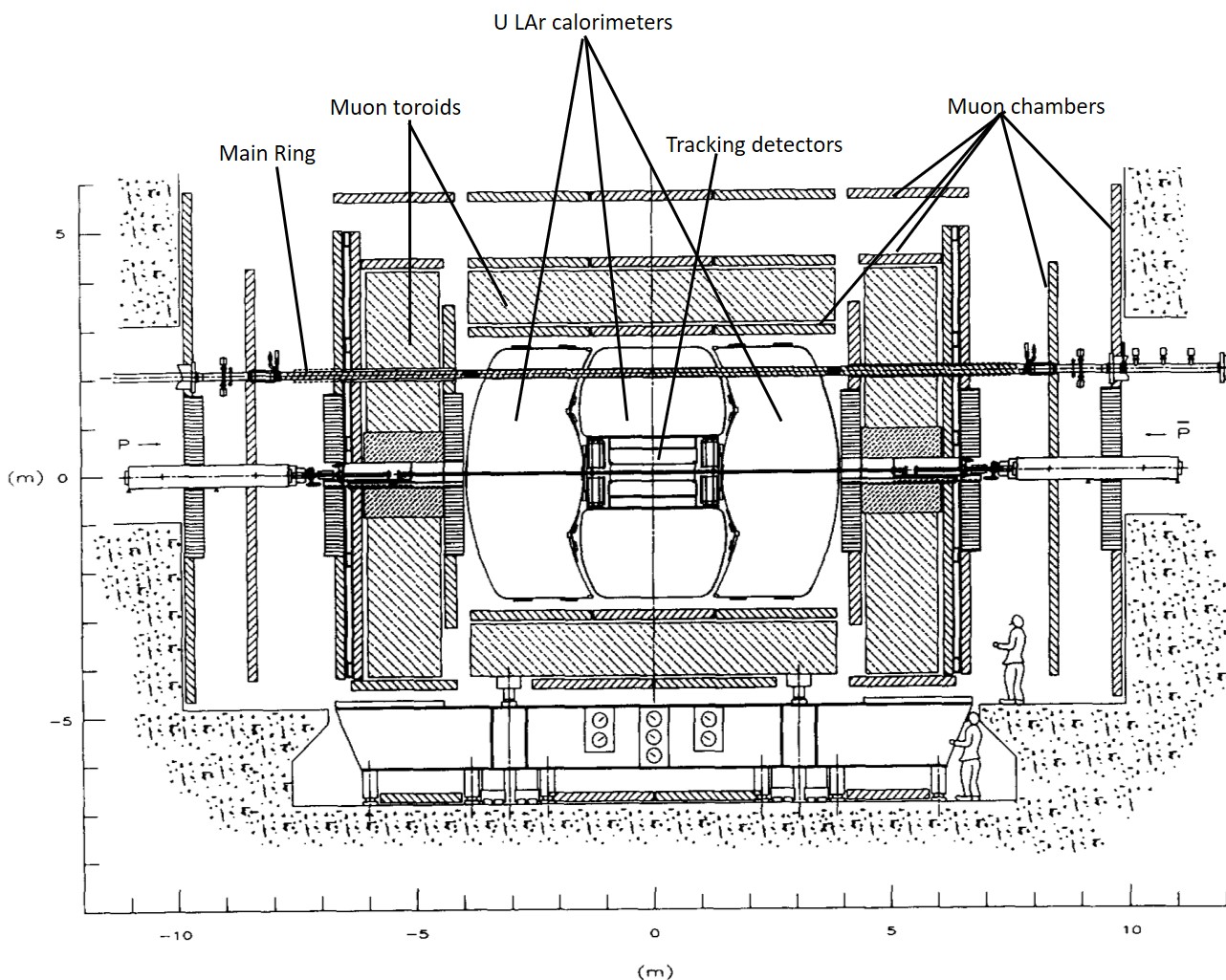}
\caption{\label{fig:d0-detector} The D0 detector, circa 1992.}
\end{figure}

In contrast, the initial DØ detector (Fig.~\ref{fig:d0-detector}) had no 
central magnetic field  
but employed a coherent liquid argon/uranium calorimeter in both barrel and 
end cap regions. The muon detection system also was moderately coherent 
and extended to very high $\eta$.

Both detectors endured the presence of the original Fermilab Main Ring, 
which was used to accelerate protons before injection in the Tevatron, 
and also to produce the anti-protons. In each intersection region an “overpass” was introduced into the Main Ring. In the case of CDF, it passed over 
the detector and was the source of background in the upper muon detectors. 
In the DØ case it passed through the uranium/liquid argon calorimeter 
and necessitated stopping data collection while the Main Ring beam was present.

Neither CDF nor DØ had been initially designed with discovery of the top quark 
in mind. However, by 1992, with operation of the UA1 and UA2 detectors at 
the S${\rm p}\bar{\rm p}$S concluded, the search for the 
top quark became a primary goal. 
In early 1994, DØ published the last lower mass limit of 131 GeV. In March, 
CDF announced evidence, limited by statistical precision, of a top quark 
with mass around 175 GeV. A year later both CDF~\cite{cdftop} and 
DØ~\cite{d0top} published their observations of the top quark, its mass and its 
production cross-section. The two experiments were not in very significant 
disagreement, although the masses quoted differed by nearly 25 GeV and the 
cross-sections by a factor of two. In the first case, DØ was higher than the 
eventually accepted value and in the latter, CDF was too high. One might note 
that it is a general characteristic of first observations that they 
benefit from upward statistical fluctuations. Ultimately, the 
combination of measurements from each experiment with a wide variety 
of decay modes resulted in a determination of the top quark mass with an 
uncertainty of less than 1 GeV, or about 0.6\%.

\begin{figure}[hbt]
\includegraphics[width=0.8\textwidth,angle=0]{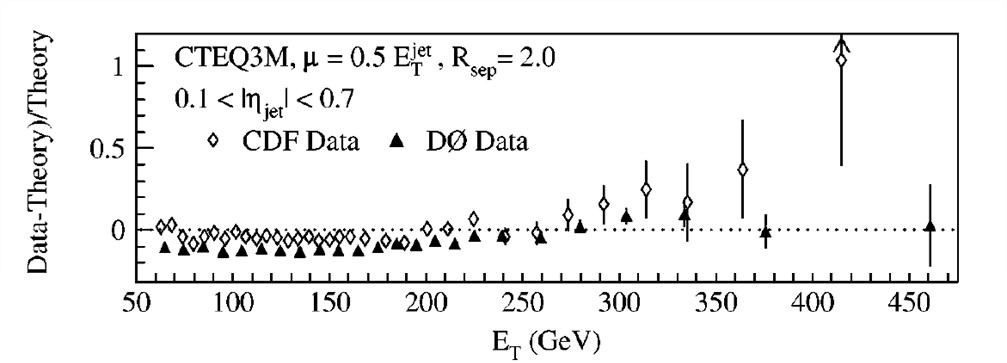}
\caption{\label{fig:jet-xs} Deviations of the jet 
production cross-sections from CDF and 
DØ relative to QCD prediction~\cite{d0-jetxs}. }
\end{figure}

Following the clear observation of jets, first in electron positron collisions, 
then in hadron-hadron collisions, the single jet production cross-section was a 
determinative test of the ability of QCD to describe the interactions. 
Deviations at high transverse momentum were anticipated in the case that the 
interacting partons had internal structure. Thus, the observation of a 
rise of the jet cross-section by CDF~\cite{cdf-jetxs} generated some excitement, 
which was muted by the contradiction and clarification of the situation by 
DØ~\cite{d0-jetxs} with higher statistical accuracy. 
This is illustrated in Fig.~\ref{fig:jet-xs}.

At the HERA Collider, both the  H1 and ZEUS experiments observed excesses of 
events at the limits of their kinematic reach~\cite{h1-lq,zeus-lq}. 
Theoretical interpretations suggested the production of leptoquarks as a 
potential source. However, the Tevatron experiments had sensitivity to 
both first and second generation leptoquarks and quickly generated 
publications setting lower mass limits in excess of 200 GeV.  
They set an important precedent by combining their data to set a lower mass limit of 242 GeV~\cite{tev-lq} on first generation leptoquarks.

\subsection{\label{sec:run2}Run 2, 1997 -- 2011}

The Tevatron Collider underwent substantial upgrades~\cite{tev-upgrade} starting in 
1997 that replaced the Main Ring with the Main Injector as the pre-accelerator 
to the Tevatron. This was built in a separate tunnel, which removed the Main Ring 
accelerator from its complicating presence in the Tevatron infrastructure. The 
Main Injector tunnel was also used to house the antiproton Recycler ring, which 
enhanced the antiproton beam intensity beyond that from the Antiproton 
Accumulator. An increase in the number of bunches and hence the collision 
frequency was a corollary.

The two experiments underwent quite radical changes, several of which were enabled 
by improvements in detector technology. DØ~\cite{d0-upgrade} recognized the potential 
of introducing a solenoidal magnetic field and accompanied it with an innovative 
scintillating fiber tracker enclosing a silicon-strip tracker including both 
barrel and disk configurations. In addition, it enhanced its muon detection system 
and upgraded the LAr calorimeter electronics to accommodate the increased 
collision frequency.  
CDF~\cite{cdf-upgrade} also completely replaced its tracking system with a new wire 
chamber enclosing a new multi-layer silicon tracker and vertex detection system. 
Improvements to the end cap calorimetry were introduced and the muon system 
coverage was improved. Both experiments made improvements to their multi-level 
trigger systems. One can note that the upgrades to each of the experiments were 
motivated both by strengthening their strong points but also by trying to address 
demonstrated weaknesses. Importantly, in each case, advances in detector 
technology from the time of the initial design, particularly electronics and data 
acquisition, were exploited. With these Run 2 upgrades, CDF and DØ became more 
similar.

\begin{figure}[hbt]
\includegraphics[width=0.8\textwidth,angle=0]{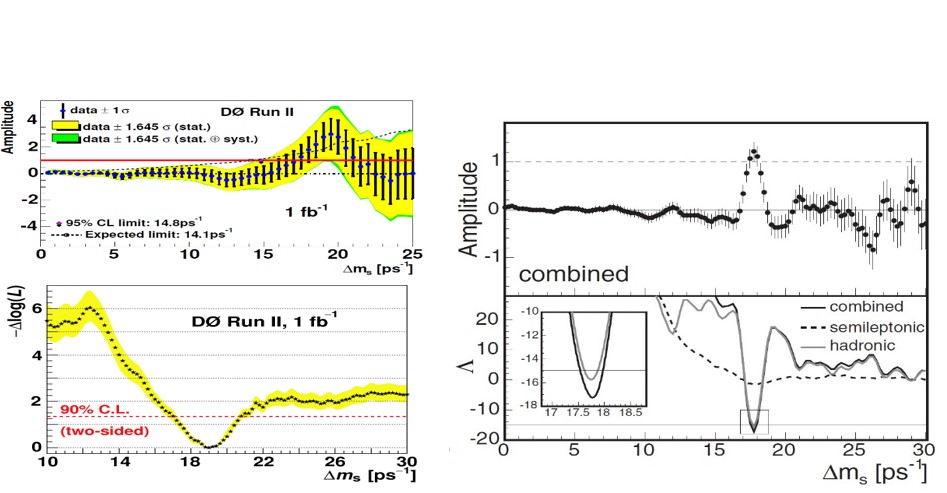}
\caption{\label{fig:bs-mixing} $B_s$ meson eigenstate mass difference: 
two sided limits from DØ on left, measurement by CDF on right. }
\end{figure}

At the time of the design of the detectors in the 1980’s, ambitions to make 
substantial contributions to $B$-physics were limited to a few people. 
However, the 
success of the CDF vertex detector in the search for the top quark, and the 
extensively improved trackers opened the door to a broad $B$-physics program. The 
LEP and SLD detectors had ceased operations without observation of $B_s$ mixing, 
which therefore became low-hanging fruit for the Tevatron experiments. CDF had a 
superior tracking and vertex system so was expected to win. However, before a 
publication from CDF, DØ recognized that its data gave a clear lower limit and 
indication of an upper limit~\cite{d0_bsmix} (see Fig.~\ref{fig:bs-mixing}) 
on the mixing frequency thus limiting, albeit relatively weakly, the central value. Therefore, they 
published. In apparent surprise, CDF produced its own more precise 
determination~\cite{cdf_bsmix-a}, which it then followed with the
improvement shown in Fig.~\ref{fig:bs-mixing} with essentially 
the same data set~\cite{cdf_bsmix-b}. A plausible interpretation is that the emergence of DØ as a 
competitor motivated the CDF teams to complete and publish their results.

This $B_s$ mixing episode was followed by a rich trove of $B$-physics 
results from both 
experiments. However, perfection was not always achieved. 
DØ published~\cite{d0-omegab} an observation of the $\Omega_b$ baryon in 2008, 
with a mass of 6165 MeV 
and uncertainty of less than 20 MeV. A year later CDF published~\cite{cdf-omegab-a} an 
observation of ostensibly the same particle but with a mass of 6054 MeV and 
precision less than 10 MeV.  CDF subsequently updated its 
result~\cite{cdf-omegab-b} to $m_{\Omega_b} = 6047.5 \pm 3.8$ MeV and LHCb found 
$m_{\Omega_b} = 6046.2 \pm 2.2$ MeV~\cite{lhcb-omegab}.  After analyzing new 
data in which the signal was not significantly present, 
DØ produced a note~\cite{do-omegab-retract} 
stating that while it had not found a mistake, it conceded that on 
the basis of the CDF and LHCb results its observation should be disregarded.

\begin{figure}[hbt]
\includegraphics[width=0.8\textwidth,angle=0]{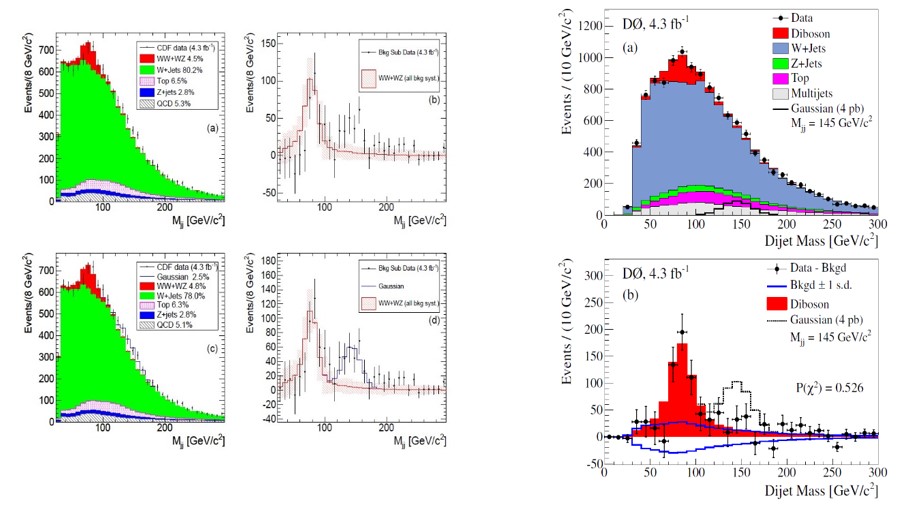}
\caption{\label{fig:dijet-resonance} Putative dijet resonance 
with mass 144 GeV: Left-CDF, Right-DØ. }
\end{figure}

A CDF 2011 study of di-jet mass distributions in conjunction with a study of $W$ 
boson plus jets seemed to show an enhancement corresponding to a mass of 144 GeV~\cite{cdf-Wjj}.
The significance corresponded to a little more than 
three standard-deviation fluctuation in the background, and a cross section of 
about 4 pb. The DØ study~\cite{d0-Wjj} closely followed the same analysis 
procedure and yielded no sign of a signal, giving a likelihood of less than $10^{-5}$
for the resonance hypothesis. The relevant distributions are illustrated in 
Fig.~\ref{fig:dijet-resonance}.

Another CDF analysis~\cite{cof-ghost} indicated the presence of so-called “ghost 
muons”, which originated at more than 1.5~cm from the interaction point. These 
constituted about 12\% of the observed muons in the study. In this instance 
DØ observed~\cite{d0-ghost} an essentially null result that such detached muons 
constituted 0.4±0.6\% of the total sample.  In each of these last three examples, 
the existence of a second experiment with the ability to perform a cross check, 
was invaluable.

\begin{figure}[hbt]
\includegraphics[width=0.8\textwidth,angle=0]{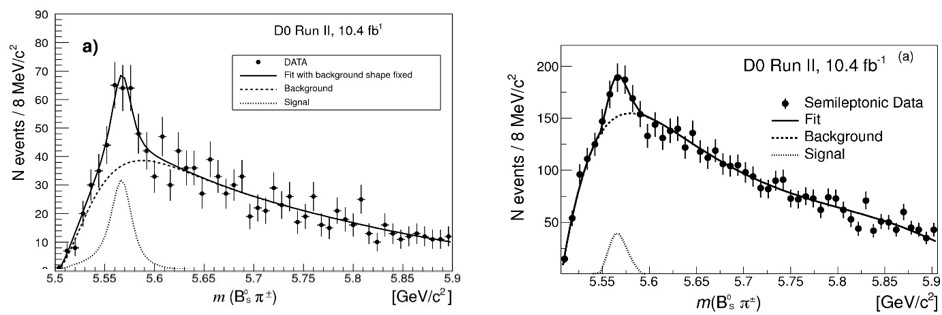}
\caption{\label{fig:Bs-pi} DØ Observation of X(5568) decay to
$B_s \pi$ with $B_s$ hadronic decay at left and semi-leptonic at right. }
\end{figure}

In the instances of disagreement discussed above, a consensus within the field 
indicated implicit, if not explicit, resolution of the disagreement. However, 
sometimes such a consensus is not easily reached. An example of continuing tension 
concerns the observation by DØ of a new exotic meson containing two 
quarks and two antiquarks, all of different flavors.  The resonance with a mass 
of 5568 MeV in $B_s \pi$ final states was seen for both $B_s$  decays to 
$J/\psi \phi$~\cite{d0-bspi-had}  and $D_s^- \mu^+ X$~\cite{d0-bspi-semilept} with a combined significance of 6.7$\sigma$ 
(see Fig.~\ref{fig:Bs-pi}). Neither LHCb~\cite{lhcb-bspi-a,lhcb-bspi-b}, 
CMS~\cite{cms-bspi}, CDF~\cite{cdf-bspi}
nor ATLAS~\cite{atlas-bspi} saw the signal and placed limits on the observation of such a state. 
The compatibility of the CDF and DØ measurements, both for $p\bar p$ production at 1.96 TeV, was
 at the 2$\sigma$ level.
It might be that the larger $\eta$ acceptance of the 
DØ muon system could explain the difference if, as suggested in the DØ data, forward muons 
were favored in the resonance decay. The LHC experiments' null 
results might be due to  
the higher partonic densities in the LHC collisions that have been seen to disrupt the 
loosely bound exotic $X(3872)$ state~\cite{d0-3872}. There are also theoretical speculations~\cite{theory-bspi} 
about how the complexity of the state could lead to these variations in the observability of 
the state. Nevertheless the issue remains unresolved according to the Particle Data Group. 

There can be cases where one experiment makes an unexpected measurement or an observation that relies on a 
particular feature of the apparatus which is not available to other experiments. An example is 
the DØ measurement~\cite{d0-dimuon} of the asymmetry between the rate for production of two $B$ 
mesons with decays that include two positively charged muons 
and the rate for two negative 
muons. The like-sign final state arises when one meson in a $B\bar B$ pair oscillates to the 
opposite flavor and both decay semi-leptonically.  The ability to reverse the polarities of both 
the central solenoid magnet and the iron toroids of the muon detector was the key to controlling 
the false asymmetries related to detector effects. DØ’s ability to check the estimates of the 
impact of backgrounds rested upon the measurement of the muon momentum in both the central 
and outer magnets.  The measured asymmetry disagreed with the SM value at the 3.9$\sigma$ level. 
CDF did not have the capability for this suppression of false asymmetries, and the result 
remains in limbo awaiting future measurements. 

\begin{figure}[hbt]
\includegraphics[width=0.8\textwidth,angle=0]{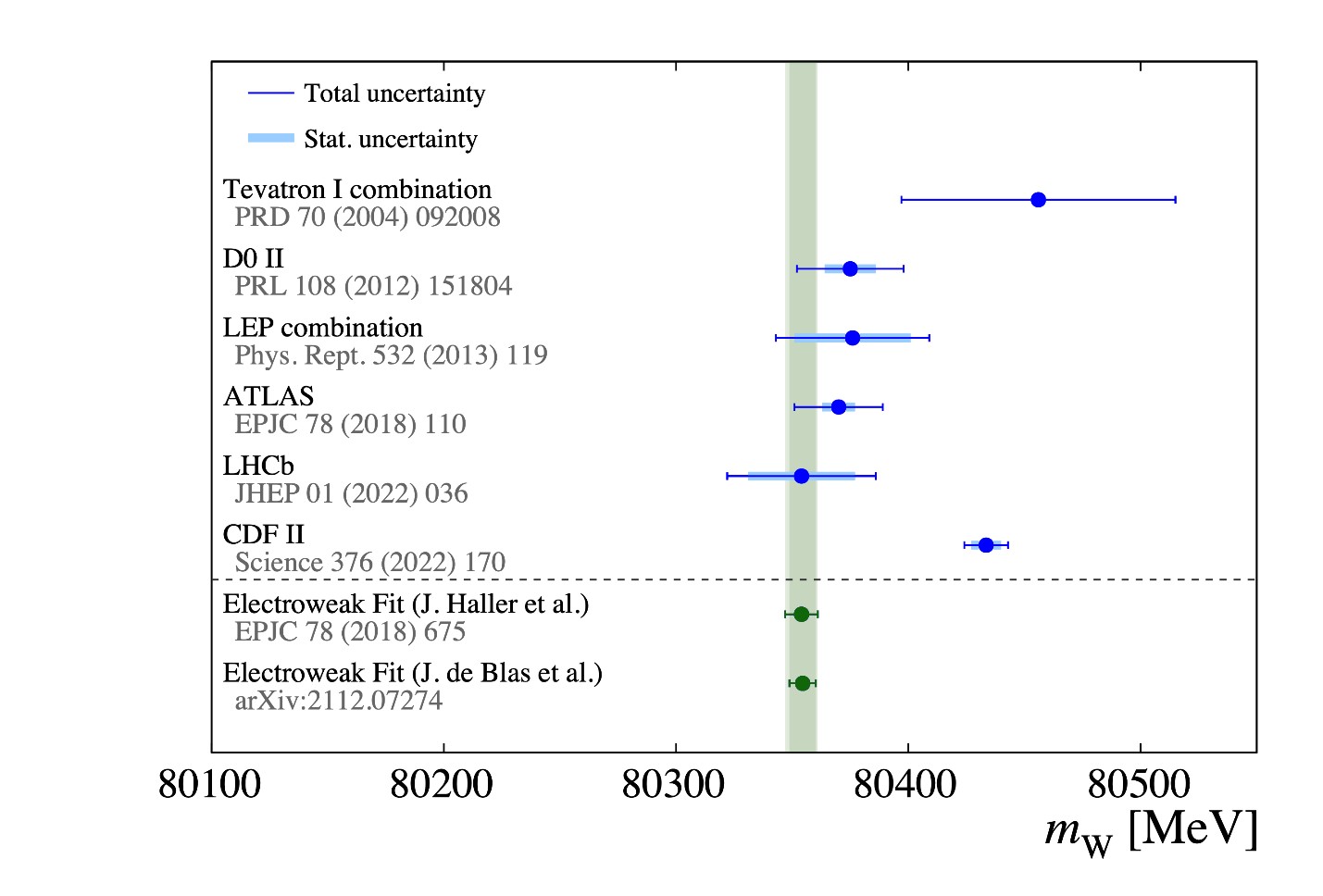}
\caption{\label{fig:world-mw} World measurements of the 
$W$ boson mass~\cite{world-mw}. }
\end{figure}

A final example of an unresolved situation concerns the recent CDF measurement 
of the $W$ boson mass.  
One of the apparent success stories for hadron colliders has been the determination of the
$W$ boson mass. UA2 demonstrated the potential and created an important legacy measurement 
from the S${\rm p}\bar{\rm p}$S  collider. The Tevatron experiments followed with a 
series of measurements with improving precision, which eclipsed the measurements from 
the electron-positron measurements at LEP. These measurements demonstrated compatibility and 
converged towards a combined precision of 16 MeV~\cite{cdfd0-mw,greatest-hits}. 
The first LHC measurements from 
the ATLAS experiment~\cite{atlas-mw}, and then the LHCb experiment presented 
results~\cite{lhcb-mw} 
with precision comparable to those of the prior Tevatron measurements and in good agreement 
with them. However, in 2022, CDF completed an analysis of all its data from the Tevatron and 
presented a new result~\cite{cdf-mw-new}, halving the uncertainty with respect to the 
best previous measurement. The big surprise was not however the uncertainty but the central value. 
This measurement is incompatible at the level of several standard deviations with 
all the previous measurements of comparable precision (see Fig.~\ref{fig:world-mw}) 
including the previous CDF measurement. The question 
is whether this is an example of a series of subsequent measurements falling on top of previous ones before 
the enlightened new measurement spoke the truth, or whether there is an elusive 
issue with the new 
measurement. The resolution of this puzzle will lie in new measurements. 	

The ability to combine results and give more incisive information 
is a further
advantage of having two experiments.
The top quark was initially observed through its pair production 
via the strong interaction. 
Single top production can occur through the electroweak interaction, 
but the more favorable  
available phase space is offset by the weakness of the coupling, 
such that the sum of the $s$-channel and $t$-channel 
production cross sections is approximately 
half that of $t\bar t$ production at 2 TeV.  
The simpler final state also means that the backgrounds to 
single top are larger. Further, the signal for
single top production is less striking than that for a 
top-antitop pair. The analysis demands a 
sophisticated approach to suppression of the background 
and multiple layers of multivariate analysis 
were used by both experiments. The results were 
successful in that the $t$-channel production was 
observed separately by each experiment, but for the 
$s$-channel, neither experiment could claim 
observation. However, when the results of both 
experiments were combined, clear observation in both 
channels could be claimed~\cite{tev-singletop} and the 
total cross-section was measured 
to be 3.3 pb, with a precision of about 0.7 pb. 
A very significant bonus from this measurement is the 
determination of the relevant CKM matrix element to be 
$|V_{tb}| > 0.92$ with 95\% CL. 

\begin{figure}[hbt]
\includegraphics[width=0.6\textwidth,angle=0]{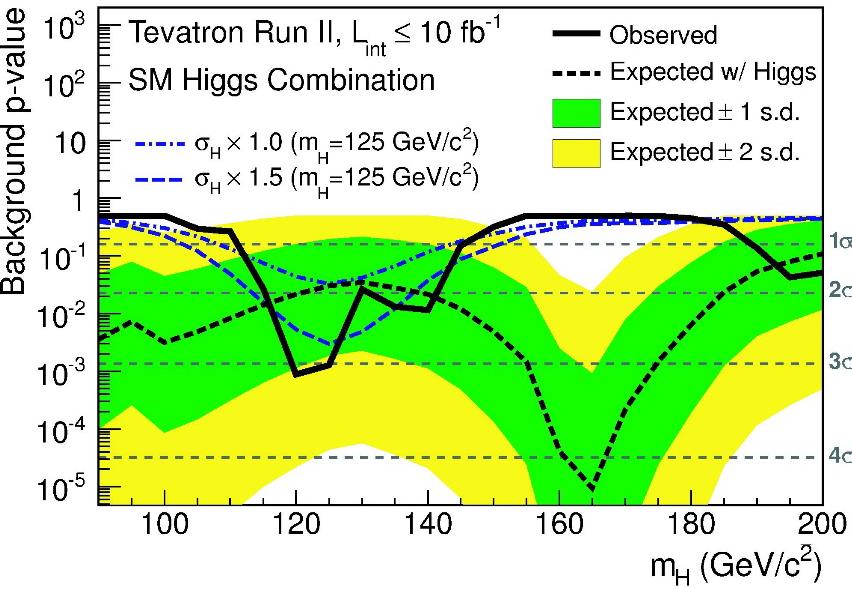}
\caption{\label{fig:tevatron-higgs} Tevatron Higgs search results.
}
\end{figure}

Much of the Tevatron’s operation in Run 2 was motivated 
by the desire 
to observe the standard model Higgs boson. It was understood 
from the outset that combination of the 
data from the two experiments would be required, given 
the expectations for integrated luminosity. 
The sensitivity to the associated $VH$ production was greater 
than that for the gluon-gluon fusion process owing the 
presence of leptons in the final state that served to suppress 
the background and provide efficient 
triggers. Leaning heavily on techniques established in the 
search for single top production, CDF 
included 15 distinct channels while DØ included 13. 
The results~\cite{tev-higgs} are illustrated in 
Fig.~\ref{fig:tevatron-higgs}. The solid line represents 
the background 
$p$-value observed in the data as a function of mass. 
The blue dash-dotted lines and dashed curves indicate expectations 
from a Higgs particle with mass of 
125 GeV and the anticipated cross section and 1.5 times that 
cross section respectively. An excess is 
indicated by a reduced $p$-value for the background hypothesis. 
Such an excess is observed in the region 110 to 140 GeV, 
which at 125 GeV is at the level of 3 standard deviations.

The results in the channel $VH(b\bar b)$  constituted evidence 
for the Higgs decay into fermions, in 
distinction with the initial observation at the LHC which was made 
using the bosonic decays $H\rightarrow \gamma \gamma$ and $H\rightarrow ZZ^*$.   
We reiterate that these results were only possible by 
combining the data and the analyses from the two experiments, which, 
in some ways, was the maximal 
return from operations at the Tevatron Collider.

\section{\label{sec:lhc}The LHC: ATLAS and CMS}

\begin{figure}[hbt]
\includegraphics[width=0.6\textwidth,angle=0]{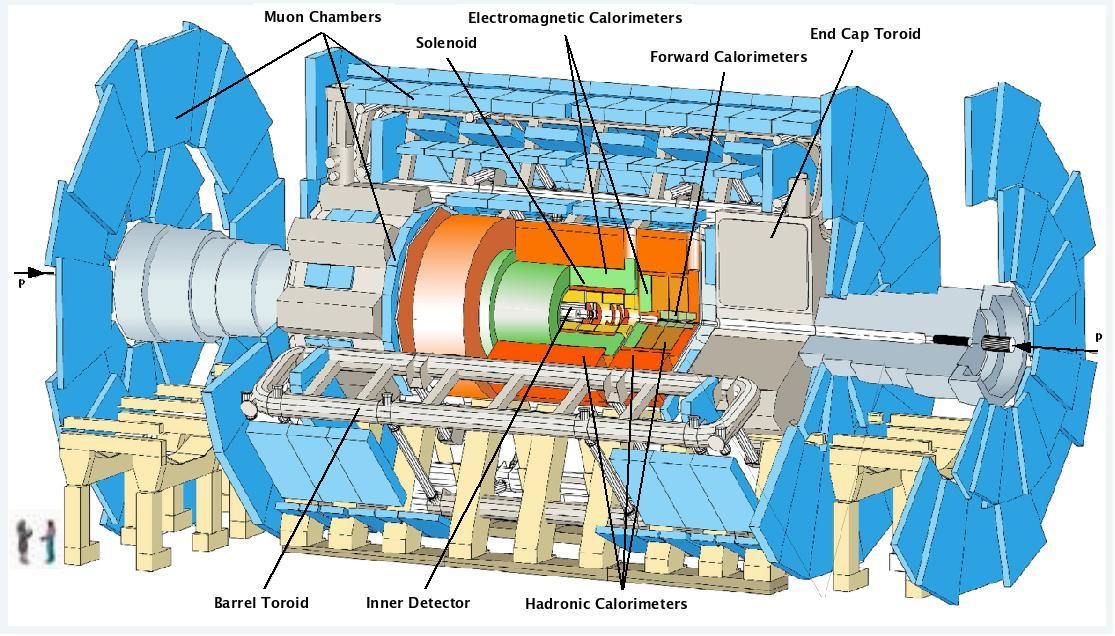}
\caption{\label{fig:atlas-det} The ATLAS detector~\cite{atlas}.
}
\end{figure}

\begin{figure}[hbt]
\includegraphics[width=0.7\textwidth,angle=0]{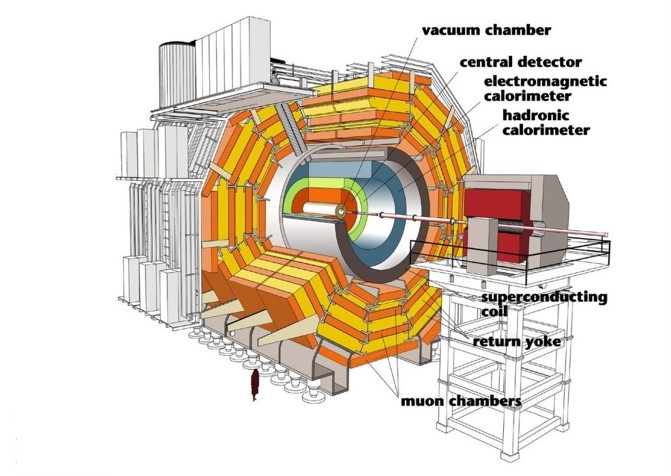}
\caption{\label{fig:cms-det} The CMS detector~\cite{cms}.
}
\end{figure}

While we are more than a decade into operations of the Large Hadron Collider, 
we expect that we have only had glimpses of its ultimate achievements. 
Nevertheless, we can draw some guidance from the experience so far. 
Sufficient resources were available {\it ab initio} for the construction of 
two viable general-purpose detectors. A challenge for 
the detectors was the anticipated luminosity and bunch structure of the collider. 
The former was such that at design luminosity an average of 20 
interactions per bunch crossing was anticipated with only 25 ns 
between each crossing. Such ferocious conditions had never been 
confronted in a collider environment and, at the time of LHC detector design, 
were not anticipated at the Tevatron.  Detector design therefore 
contained significant reliance on simulation and resulted in 
quite different approaches for the two general purpose detectors, 
ATLAS)~\cite{atlas} (Fig.~\ref{fig:atlas-det}), 
and the CMS~\cite{cms} (Fig.~\ref{fig:cms-det}). In fact each of the emerging 
detectors followed the merging of two other competitive proposals.

ATLAS emphasized external large air-core toroids which provided the 
detection and high precision measurement of muons over a large 
range in rapidity. Because the calorimeter shields the muon system 
from hadronic interaction debris, it represents an approach to risk 
mitigation. The resulting tracking system is necessarily modest 
in volume and the solenoidal magnet which supports the hadron 
electron momentum measurement is quite thin. The tracking system 
emphasized the number of hits per track by using thin straw tube 
detectors, which also reduced the potential impact of broken wires 
by limiting the loss to the single wire. The tracking included the 
capability to require a signal consistent with transition radiation 
which is only generated by electrons. This was a further potential mitigation 
of the hadronic track pileup from the multiple interactions. 
The thin solenoid minimizes the degradation of the  
electron and photon components in the showers before their detection in 
high quality liquid argon calorimeters. Here one sees that despite 
the expected potential difficulties associated with multiple 
interactions in the inner detector, the need to detect photons 
from the decay of the Higgs boson played a seminal role in 
benchmarking the design. The electronics was such that, 
in principle, only those transition radiation hits could be used thus, 
again, suppressing hadronic track background. The innermost tracking, 
silicon strips and pixels served to identify the primary and secondary vertices.

In contrast to ATLAS, CMS chose a large high-field solenoid as the base 
component of the design. It is large enough to contain the calorimeter 
whose electromagnetic sections were constructed of 
lead tungstate crystals. 
Obtaining good electromagnetic energy resolution for one crystal is 
straightforward. For a system of thousands of crystals advanced control 
of temperature and humidity for the whole system is the challenge. 
The muon detection system consists of multilayers of detectors
interspersed with the iron magnetic flux return and 
so is quite compact compared to the ATLAS air-core toroids. 
CMS chose to use silicon detectors for all of its tracker. 
This innovative choice had no precedent. This choice emphasizes 
hit precision rather than the numbers of hits on a track.  
It also depends on the precision of the track-hits to offset 
the significant multiple scattering in each detector layer.  
As with ATLAS, pixel detectors in the layers closest to the 
interaction point were key to the vertex measurements.

For both experiments, the limitations of the data acquisition capabilities relative 
to the numbers of interactions and the sizes of the data sets 
from each crossing necessitated strong constraints on the 
fraction of events accepted. In turn that capability suppression 
necessary dictated multiple levels within the trigger systems. 

It is remarkable how, sub-system by sub-system, the two experiments 
made complementary choices. What is then further remarkable is 
that in terms of physics performance, the two experiments are 
largely matching each other.

\begin{figure}[hbt]
\includegraphics[width=0.8\textwidth,angle=0]{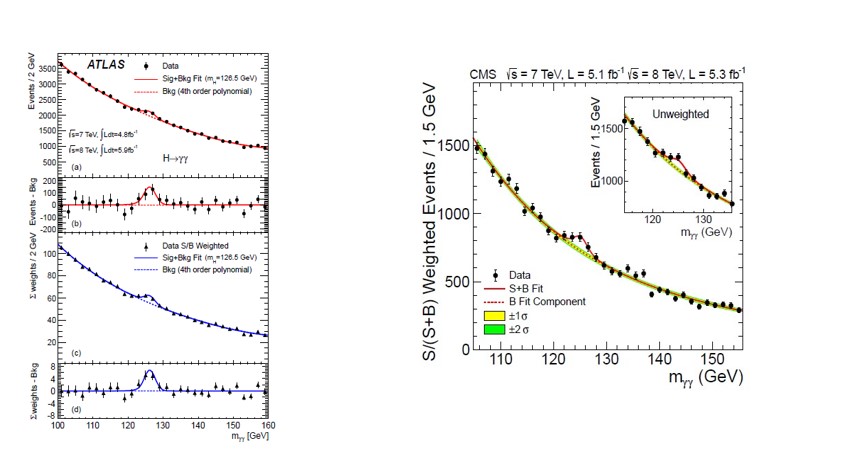}
\caption{\label{fig:lhc-higgs} Higgs diphoton signals from ATLAS (left) and CMS (right).
}
\end{figure}

Of course, the most prominent result from the LHC was the 
observation of the Higgs boson~\cite{atlas-higgs,cms-higgs} through its decay 
into two photons (as well as into four leptons from real and virtual $Z$ bosons). 
As we discussed above, the projected capability to make this 
measurement was a strong discriminator in the detector designs. 
The signal is an enhancement in the mass spectrum constructed for 
two photons. Of course, much of the spectrum results from backgrounds 
so that the signal is a modest enhancement visible but not dominant. 
The spectra from the two experiments are shown in Fig.~\ref{fig:lhc-higgs}. 
The similarity of the spectra in the two experiments and of the 
significances quoted demonstrate how the performances of the 
two detector designs are very similar.
             
\section{\label{sec:summary}Summary}

The benefit of having two experiments arises in several different ways. 
If the specific technological character of a subsystem of the two detectors 
is different (for example crystal calorimetry with excellent energy 
resolution compared with a sampling calorimeter with lower resolution 
but higher segmentation) the systematic uncertainties of a measurement 
can be quite different rendering the combined result more robust. 
If the two experiments’ subsystems have some orthogonality, 
the combined understanding of the physics can be amplified. 
For example, if one detector has particle identification capability 
as opposed to precision timing measurements for its track or 
calorimetric signals or different orientations of magnetic fields, 
the sum of the physics that can be addressed by the two experiments is enlarged.
 
There is another axis along which the utility of two experiments can be realized. 
The simple existence of two independent sets of physicists allows for different solutions to common problems that may be more important even than specific 
hardware differences. The collaborations may invent different algorithms 
for such problems as pileup mitigation or jet substructure characterization 
or develop different methods for performing position or energy calibrations. 
Once such differences become public, they of course will enter the common 
toolkit used by the other experiment and subsequent experiments.  
But the cross fertilization of ideas from the competing groups of 
physicists can enhance the overall effectiveness of the program. 
This independent attack on major analysis techniques may well 
be the greatest benefit of having more than one experiment at the collider.

In this note we have discussed specific examples in which having two 
detectors made improvements to the physics results of a collider program.  

Even within a collaboration that produces a new result, there are often 
those who express doubts about its validity. Thus, even with just one 
experiment, there are self-correcting forces in play.  However, the 
tendency for modern collider collaborations to seek a common voice 
on their results means that this sort of regulation is not fully effective. 
We have seen several instances where one experiment announced a new discovery, 
but the other experiment performed a similar analysis that 
contradicted it and effectively nullified the discovery.  
There is a continuum of such cases, ranging from a putative new 
discovery outside the standard model paradigm to quantitative results 
on a cross section or kinematic parameter that are contradicted 
by the second measurement.  In such cases, the reliability of the 
overall collider program is clearly enhanced.  

We presented several cases in which one experiment presents a result 
with significant consequences for our understanding of particle physics 
where the second experiment did not have the capability to confirm a new result. 
Even though the new result is not negated, the inability to corroborate 
served to flag the measurement as tentative, in need of confirmation 
by some future experiment.

We have seen how the competition between two experiments explicitly 
accelerated the appearance of key results thus leading to a higher quality program. 

There are extremely important cases in which neither experiment by 
itself is capable of measurement of some physical observable with 
convincing significance, but through combination of the two, the 
observable is definitively established. This coherent effect of 
combining the two experiment’s results is routinely useful in 
improving the precision of an observable or in extending the kinematic 
range of a measurement.  And in these confirmatory cases, the 
agreement of the measurements by the two experiments serves to validate both.

Although our argument has been drawn primarily from the experience at 
hadron colliders, an analysis for the International Linear Collider 
reached a similar conclusion~\cite{brau}, despite the fact that in this case 
the two detectors cannot take data simultaneously.

In all the cases we have examined, the benefit of mounting more than 
one experiment has been large.   Given that any one experiment has been 
seen to make mistakes some fraction of the time, there is always the 
nagging possibility that, if there is but one experiment, any 
particular result is incorrect, and the impact of the whole collider 
program is compromised.

\begin{acknowledgments}
We express our appreciation to Dmitri Denisov for his comments during the development of the talk for the workshop and to Jim Brau for bringing the similar study for a linear collider to our attention.   HEM would also like to thank the organizers of the workshop, in particular Pawel Nadal-Turonski, whose encouragement stimulated this work. This material is based upon work supported by the U.S. Department of Energy, Office of Science, Office of Nuclear Physics under contract DE-AC05-06OR23177.
\end{acknowledgments}


\begin{thebibliography}{0}%
\makeatletter
\providecommand \@ifxundefined [1]{%
 \@ifx{#1\undefined}
}%
\providecommand \@ifnum [1]{%
 \ifnum #1\expandafter \@firstoftwo
 \else \expandafter \@secondoftwo
 \fi
}%
\providecommand \@ifx [1]{%
 \ifx #1\expandafter \@firstoftwo
 \else \expandafter \@secondoftwo
 \fi
}%
\providecommand \natexlab [1]{#1}%
\providecommand \enquote  [1]{``#1''}%
\providecommand \bibnamefont  [1]{#1}%
\providecommand \bibfnamefont [1]{#1}%
\providecommand \citenamefont [1]{#1}%
\providecommand \href@noop [0]{\@secondoftwo}%
\providecommand \href [0]{\begingroup \@sanitize@url \@href}%
\providecommand \@href[1]{\@@startlink{#1}\@@href}%
\providecommand \@@href[1]{\endgroup#1\@@endlink}%
\providecommand \@sanitize@url [0]{\catcode `\\12\catcode `\$12\catcode
  `\&12\catcode `\#12\catcode `\^12\catcode `\_12\catcode `\%12\relax}%
\providecommand \@@startlink[1]{}%
\providecommand \@@endlink[0]{}%
\providecommand \url  [0]{\begingroup\@sanitize@url \@url }%
\providecommand \@url [1]{\endgroup\@href {#1}{\urlprefix }}%
\providecommand \urlprefix  [0]{URL }%
\providecommand \Eprint [0]{\href }%
\providecommand \doibase [0]{https://doi.org/}%
\providecommand \selectlanguage [0]{\@gobble}%
\providecommand \bibinfo  [0]{\@secondoftwo}%
\providecommand \bibfield  [0]{\@secondoftwo}%
\providecommand \translation [1]{[#1]}%
\providecommand \BibitemOpen [0]{}%
\providecommand \bibitemStop [0]{}%
\providecommand \bibitemNoStop [0]{.\EOS\space}%
\providecommand \EOS [0]{\spacefactor3000\relax}%
\providecommand \BibitemShut  [1]{\csname bibitem#1\endcsname}%
\let\auto@bib@innerbib\@empty
\end{thebibliography}%


\begin{thebibliography}{99}

\bibitem{eic} 
The Electron Ion Collider: https://www.bnl.gov/eic/machine.php.

\bibitem{eic-wg} 
EIC 2nd Detector WG: https://indico.bnl.gov/event/17693/contributions/70922
\newline/attachments/44976/75872/EICUG\_2ndDetector\_Incubator.pdf.

\bibitem{eic-wkshp} 
EIC Users Group 2nd Detector Meeting: https://indico.bnl.gov/event/17693/.
 
 \bibitem{spps}
 The Super Proton-antiproton Collider: https://en.wikipedia.org/wiki/Super\_Proton-Antiproton\_Synchrotron. 

\bibitem{tevatron}
The Tevatron Collider: https://en.wikipedia.org/wiki/Tevatron. 

\bibitem{cdf}
The CDF Detector: https://en.wikipedia.org/wiki/Collider\_Detector\_at\_Fermilab. 

\bibitem{d0}
The DØ Detector: https://en.wikipedia.org/wiki/ DØ\_experiment.

\bibitem{cdftop}
F. Abe {\it et al.} (CDF Collaboration), Phys. Rev. Lett. {\bf 74}, 2626 (1995).

\bibitem{d0top} 
S. Abachi {\it et al.} (DØ Collaboration), Phys. Rev. Lett. {\bf 74}, 2632 (1995).

\bibitem{cdf-jetxs}
F. Abe {\it et al.} (CDF Collaboration), Phys. Rev. Lett. {\bf 77}, 438 (1996).

\bibitem{d0-jetxs}
B. Abbott {\it et al.} (DØ Collaboration), Phys. Rev. Lett. {\bf 82}, 2451 (1999).

\bibitem{h1-lq} 
C. Adloff {\it et al.} (H1 Collaboration), Zeitschrift für Physik C{\bf 74}, 191 (1997).

\bibitem{zeus-lq}
J. Breitweg {\it et al.} (ZEUS Collaboration), Zeitschrift für Physik C{\bf 74}, 207 (1997).

\bibitem{tev-lq}
CDF and DØ Collaborations, arXiV:9810015 (hep-ex) (1998).  

\bibitem{tev-upgrade}
Stephen D. Holmes and Vladimir D. Shiltsev, Annu. Rev. Nucl. Part. Sci. {\bf 63}, 435 (2013).

\bibitem{d0-upgrade}
V. M. Abazov {\it et al.} (DØ Collaboration), Nucl. Instrum. Meth. A {\bf 565}, 463 (2006).

\bibitem{cdf-upgrade}
T. LeCompte and H. T. Diehl, The CDF \& DØ Upgrades for Run II: Annu. Rev. Nucl. Part. Sci. {\bf 50}, 71 (2000).

\bibitem{d0_bsmix}
V. M. Abazov {\it et al.} (DØ Collaboration), Phys. Rev. Lett. {\bf 97}, 021802 (2006).

\bibitem{cdf_bsmix-a}
A. Abulencia {\it et al.} (CDF Collaboration), Phys. Rev. Lett. {\bf 97}, 062003 (2006).

\bibitem{cdf_bsmix-b}
A. Abulencia {\it et al.} (CDF Collaboration), Phys. Rev. Lett. {\bf 97}, 242003 (2006). 

\bibitem{d0-omegab}
V.M. Abazov {\it et al.} (DØ Collaboration), Phys. Rev. Lett. {\bf 101}, 232002 (2008).

\bibitem{cdf-omegab-a}
T. Aaltonen {\it et al.} (CDF Collaboration), Phys. Rev. D{\bf 80}, 072003 (2009). 

\bibitem{cdf-omegab-b}
T. Aaltonen {\it et al.} (CDF Collaboration), Phys. Rev. D{\bf 89}, 072014 (2014).

\bibitem{lhcb-omegab}
R. Aaij {\it et al.} (LHCb Collaboration), Phys. Rev. Letters {\bf 110}, 182001 (2013).

\bibitem{do-omegab-retract}
DØ Collaboration,   ``Update on observations using the full Run2 data set”,
 \newline https://d0.fnal.gov/Run2Physics/WWW/results/final/B/B08G/  (2015).

\bibitem{cdf-Wjj}
T. Aaltonen {\it et al.} (CDF Collaboration), Phys. Rev. Lett. {\bf 106}, 171801 (2011).

\bibitem{d0-Wjj}
V.M. Abazov {\it et al.} (DØ Collaboration), Phys. Rev. Lett. {\bf 107}, 011804 (2011).

\bibitem{cof-ghost}
T. Aaltonen {\it et al.} (CDF Collaboration), arXiV:0810.5357 [hep-ex] (2008).

\bibitem{d0-ghost}
DØ Collaboration, DØ Note 5905-CONF (2009).

\bibitem{d0-bspi-had}
V.M. Abazov {\it et al.} (DØ Collaboration), Phys. Rev. Lett. {\bf 117}, 022003 (2016). 

\bibitem{d0-bspi-semilept}
V.M. Abazov {\it et al.} (DØ Collaboration), Phys. Rev. D {\bf 97}, 092504 (2018).

\bibitem{lhcb-bspi-a}
R. Aaij, {\it et al.} (LHCb Collaboration), Phys. Rev. Lett. {\bf 117}, 152003 (2016). 

\bibitem{lhcb-bspi-b}
R. Aaij, {\it et al.} (LHCb Collaboration), Phys. Rev. Lett. {\bf 118},  109904 (2017).

\bibitem{cms-bspi}
M. Sirunyan, {\it et al.} (CMS Collaboration), Phys. Rev. Lett. {\bf 120}, 202005 (2018).

\bibitem{cdf-bspi}
T. Aaltonen, {\it et al.} (CDF Collaboration), Phys. Rev. Lett. {\bf 120}, 202006 (2018).

\bibitem{atlas-bspi}
M. Aaboud, {\it et al.} (ATLAS Collaboration), Phys. Rev. Lett. {\bf 120}, 202007 (2018).

\bibitem{d0-3872}
V.M. Abazov {\it et al.} (DØ Collaboration), Phys. Rev. D{\bf 102}, 072005 (2020).

\bibitem{theory-bspi}
Hong-wei Ke and Xue-Qian Li, Phys. Lett. B{\bf 785}, 301 (2018).

\bibitem{d0-dimuon}
V.M. Abazov {\it et al.} (DØ Collaboration), Phys. Rev. D{\bf 89}, 012002 (2014).

\bibitem{cdfd0-mw}
T. Aaltonen {\it et al.} (CDF \& DØ Collaboration), Phys. Rev. D{\bf 88}, 05218 (2013).

\bibitem{greatest-hits}
Dmitri Denisov and Costas Vellidis, Rep. Prog. Phys. {\bf 85}, 116201 (2022).

\bibitem{atlas-mw}
M. Aaboud {\it et al.} (ATLAS Collaboration), Eur. Phys. J. C{\bf 78}, 110 (2018).

\bibitem{lhcb-mw}
R. Aaij {\it et al.} (LHCb Collaboration), JHEP 01  036 (2022).

\bibitem{cdf-mw-new}
T. Aaltonen {\it et al.} (CDF Collaboration), Science {\bf 376}, 170 (2022).

\bibitem{world-mw}
``World W Mass Measurements, 
https://cerncourier.com/a/cdf-sets-w-mass-against-the-standard-model/
\newline CERN Courier, April 2022.

\bibitem{tev-singletop}
T. Aaltonen {\it et al.} (CDF \& DØ Collaboration), Phys. Rev. Lett. {\bf 115}, 152003 (2015).

\bibitem{tev-higgs}
T. Aaltonen {\it et al.} (CDF \& DØ Collaboration), Phys. Rev. D{\bf 88}, 052014 (2013) .

\bibitem{atlas}
The ATLAS Experiment:  https://atlas.cern/Discover/Detector, 
https://en.wikipedia.org/wiki/ATLAS\_experiment.

\bibitem{cms}
The CMS Experiment: https://cms.cern/, 
https://en.wikipedia.org/wiki/Compact\_Muon\_Solenoid.

\bibitem{atlas-higgs}
G. Aad {\it et al.} (ATLAS Collaboration), Phys. Lett. B{\bf 716} (2012) 1. 

\bibitem{cms-higgs}
S. Chatrchyan {\it et al.} (CMS Collaboration), Phys. Lett. B{\bf 716} (2012) 30.

\bibitem{brau}
James Brau, Tsunehiko Omori and Ronald Settles, in Proceedings of the 2005 International Collider Physics and Detector Workshop and 2nd ILC Accelerator Workshop, SLAC R-798, ed. N. Graf (2006).



\end{thebibliography}

\end{document}